# Omori Law. To the 100th anniversary of death of the famous Japanese seismologist

A. Guglielmi[1], B. Klain[2], A. Zavyalov[1], O. Zotov[1,2]

[1] *Schmidt Institute of Physics of the Earth, Russian Academy of Sciences, Moscow, Russia, guglielmi@mail.ru, zavyalov@ifz.ru*
[2] *Borok Geophysical Observatory, the Branch of Schmidt Institute of Physics of the Earth, Russian Academy of Sciences, Borok, Yaroslavl Region, Russia, klb314@mail.ru, ozotov@inbox.ru*

**Abstract**

One hundred years ago, Fusakichi Omori died. Our paper is dedicated to his memory. Omori made an outstanding contribution to the physics of earthquakes. In 1894 he formulated the law of aftershock evolution. Omori's Law states that after the main shock of an earthquake, the frequency of aftershocks decreases hyperbolically with time. In this paper, we briefly describe one of the directions of modern aftershock research.

## 1. Introduction

One hundred years ago, on November 8, 1923, Fusakichi Omori, the outstanding seismologist who discovered the law of aftershock evolution, the first law of earthquake physics, passed away. The law states that after a strong earthquake, the frequency of aftershocks, i.e. earthquakes following the main shock, on average, hyperbolically decreases with time:

$$n(t) = \frac{k}{c+t}. \tag{1}$$

Here $n(t)$ is the frequency of aftershocks averaged over physically infinitely small time intervals $t \geq 0$, the parameter $c > 0$ is determined by the initial condition $n(0)$, and the parameter $k > 0$ characterizes a particular event [1]. It is worth noting that Omori formulated the law that bears his name when he was 26 years old [2].

Recall that Fxakichi Omori was born in 1868 [3]. He received his education at the Imperial University of Tokyo. His teachers were John Milne and Seikei Sekiya. Omori became the professor of seismology at Tokyo Imperial University in 1896. He traveled a lot, visited the scientific centers of Europe and America. September 1, 1923 the Great Kanto Earthquake destroyed Tokyo and killed hundreds of thousands of people. Fusakichi Omori learned of the disaster while attending a conference in Australia. He immediately left for his homeland.

During the sea passage, his health deteriorated sharply, and he died shortly after his return to Tokyo at the age of 55 [4].

The significance of Omori's discovery lies in the fact that the law of aftershock evolution is still in demand, arouses keen interest, and is actively used in experimental and theoretical studies of earthquakes (e.g., see [5–10]). In this paper, dedicated to the memory of Fusakichi Omori, we will briefly describe one of the trends in the modern development of his idea. More details can be found in the review papers [9, 10].

## 2. Evolution equation

Let us pay attention to the fact that the hyperbola is the resolvent of a simple differential equation with a quadratic non-linearity. This fact allows us to reformulate the Omori law (1), writing it in the form of the aftershock evolution equation

$$\frac{dn}{dt} + \sigma n^2 = 0. \qquad (2)$$

Here, $\sigma$ is the deactivation coefficient of the earthquake source "cooling down" after the main shock [11]. Indeed, at $\sigma = \text{const}$ the solution of equation (2) coincides with the algebraic formula (1) if we set $c = \tau_0 / \sigma$, $k = 1/\sigma$, $\tau_0 = 1/n(0)$.

Writing Omori's law in differential form has a number of advantages. In the evolution equation (2) there is no rigid restriction $\sigma = \text{const}$. Accordingly, the solution of the equation takes a more general form

$$n = 1/(\tau_0 + \tau), \qquad (3)$$

where

$$\tau = \int_0^t \sigma(t') dt'. \qquad (4)$$

Formula (3), like the classical formula (1), expresses the hyperbolic dependence of the frequency of aftershocks on time, and at the same time, it takes into account that after the mainshock time in the source, figuratively speaking, flows unevenly.

Further, equation (2) suggests to us possible ways of generalizing the law of aftershock evolution. We will consider this aspect of the problem in the next section of the paper. Finally, equation (2) makes it possible to formulate and relatively easily solve the inverse problem of the earthquake source.

The inverse problem is to calculate the deactivation coefficient from observational data on the aftershock frequency. Let's rewrite (2) in the form $\sigma = dg/dt$, where the auxiliary function $g = 1/n$ is introduced. We have obtained a formal solution to the inverse problem, but the solution is unstable due to fluctuations in the original function $n(t)$. It is necessary to perform the regularization, which in this case consists in replacing $g \to \langle g \rangle$, where the angle brackets denote the operation of smoothing the auxiliary function. As a result, the solution takes the form

$$\sigma = \frac{d}{dt}\langle g \rangle. \tag{5}$$

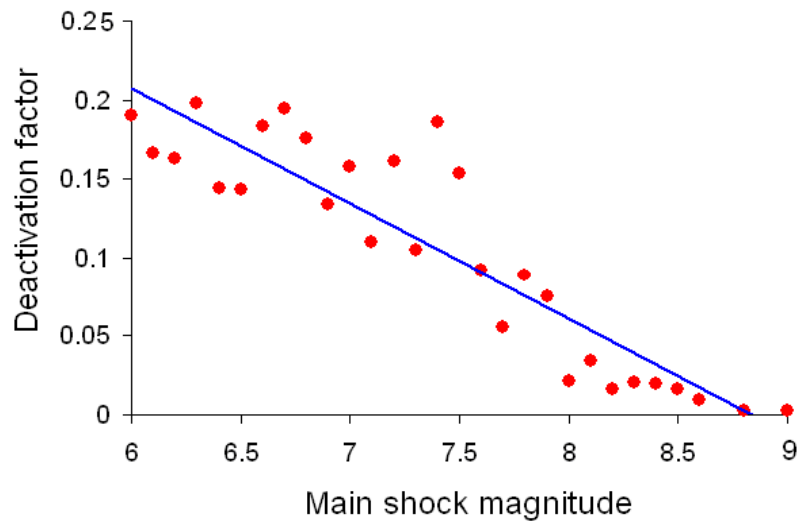

Dependence of the source deactivation coefficient on the magnitude of the mainshock [12].

It follows from theoretical considerations that the source deactivation coefficient is the smaller, the greater the magnitude of the mainshock. The solution of the inverse problem presented in the figure confirms the prediction of the theory.

## 3. Discussion

It is quite natural to add the free term $f(t)$ to the right side of equation (2). We get a non-homogeneous differential equation. It models the impact of triggers on the earthquake source.

There is a vast literature devoted to endogenous and exogenous triggers. The impact of triggers on the source, generally speaking, leads to a deviation from the hyperbolic Omori law. We confine ourselves to mentioning two characteristic triggers generated by the main shock of the earthquake. One of them is periodic ($f \propto \sin(\omega t)$), the other is pulsed ($f \propto \delta(t)$).

The periodic trigger occurs as follows. The mainshock of an earthquake excites the spheroidal and toroidal oscillations of the Earth. A modulation of the aftershock activity at a resonant frequency of 0.309 mHz of spheroidal oscillations of $_0S_2$ was found [13]. An impulsive trigger in the form of the round-the-world seismic echo excites a strong aftershock approximately 3 hours after the mainshock [14].

Faraoni drew attention to the fact that equation (2) can be represented as the Lagrange equation and proposed a bold extrapolation of the phenomenological theory of aftershocks [15]. In [16], the Faraoni Lagrangian was modified and the logistic equation was derived

$$\frac{dn}{dt} = n(\gamma - \sigma n) \tag{6}$$

to describe the evolution of aftershocks. Here $\gamma$ is the second phenomenological parameter of the theory. With the help of the logistic equation, a phase portrait of a dynamical system was constructed that simulates the evolution of aftershocks.

One more step towards generalizations can be made by adding a diffusion term to equation (6):

$$\frac{\partial n}{\partial t} = n(\gamma - \sigma n) + D\frac{\partial^2 n}{\partial x^2}. \tag{7}$$

Here $D$ is the third phenomenological parameter. By design, the nonlinear diffusion equation (7) describes the spatiotemporal evolution of aftershocks [17].

## 4. Conclusion

In conclusion, we would like to point out the ontological significance of the study of aftershocks, brilliantly begun by Fusakichi Omori 130 years ago. Modern research has enriched the physics of earthquakes with the concept of proper time and made it possible to refine the definition of the mainshock.

After the main shock, the source proper time is determined by formula (4). The unevenness of the flow of time is associated with the non-stationarity of the parameters of the geological environment in the source, and the non-stationarity is explained by the processes of relaxation of rocks to a new state of equilibrium after the mainshock. The concept of proper time can be generalized and be useful in the study of global seismic activity [18].

The ontology of earthquake physics contains a terminological base in which the definition of the mainshock is essentially based on the concept of aftershocks. Usually, the mainshock is called an earthquake, the magnitude of which $M_{msh}$ exceeds the maximum magnitude $M_{ash}$ of the aftershocks by at least a value equal to one. The inequality $M_{msh} - M_{ash} > 1$ is known as Bath's law [19]. It is quite obvious that this definition should be supplemented with a similar restriction on the magnitude of foreshocks $M_{fsh}$. To do this, one can use the recently discovered law $M_{msh} - M_{fsh} > 0.5$ [10, 20]. In this case, we obtain a fairly rigorous and practically convenient definition of the mainshock.